\documentclass{article}
\usepackage[english]{babel}
\usepackage[utf8]{inputenc}
\usepackage[table]{xcolor}
\usepackage{tabularx}
\usepackage{caption}
\usepackage{subcaption}
\usepackage{geometry}
\usepackage{cite}
\usepackage{booktabs}
\usepackage{tcolorbox}
\usepackage{floatrow}
\usepackage{multirow}
\usepackage{longtable}
\usepackage{url}
\title{Agile Effort Estimation: Have We Solved the
Problem Yet? Insights From A Second Replication Study \\{\large (GPT2SP Replication Report)}}
\author{Vali~Tawosi,~
        Rebecca~Moussa,~
        Federica~Sarro~\\
        \small{\{\textit{vali.tawosi, rebecca.moussa.18, f.sarro}\}}\small{\textit{@ucl.ac.uk}}~} 
\date{}

\begin{document}

\maketitle


\section*{}
Fu and Tantithamthavorn \cite{fu2022gpt2sp} have recently proposed GPT2SP, a Transformer-based deep learning model for SP estimation of user stories.

They empirically evaluated the performance of GPT2SP on a dataset shared by Choetkiertikul et al. \cite{deep2018} including 16 projects with a total of 23,313 issues. They benchmarked GPT2SP against two baselines (namely the naive Mean and Median estimators) and the method previously proposed by Choetkiertikul et al. \cite{deep2018} (which we will refer to as DL2SP from now on) for both within- and cross-project estimation scenarios, and evaluated the extent to which each components of GPT2SP contribute towards the accuracy of the SP estimates.

Their results show that GPT2SP outperforms DL2SP with a 6\%-47\% improvement over MAE for the within-project scenario and a 3\%-46\% improvement for the cross-project scenarios. 

However, when we attempted to use the GPT2SP source code made available by Fu and Tantithamthavorn \cite{fu2022gpt2sp} to reproduce their experiments, we found a bug in the computation of the Mean Absolute Error (MAE), which may have inflated the GPT2SP's accuracy reported in their work. Therefore, we had issued a pull request to fix such a bug, which has been accepted and merged into their repository at \url{https://github.com/awsm-research/gpt2sp/pull/2}.

In this report, we describe the results we achieved by using the fixed version of GPT2SP to replicate the experiments conducted in the original paper for RQ1 and RQ2.
Following the original study, we analyse the results considering the Medan Absolute Error (MAE) of the estimation methods over all issues in each project, but we also report the Median Absolute Error (MdAE) and the Standard accuracy (SA) for completeness.

For completeness, we also report the results of three other approaches used as a benchmark to assess and compare GPT2SP's performance, namely DL2SP, Mean, and Median estimators. These results are taken from our previous replication study of DL2SP \cite{tawosi2022deep}.

\textbf{RQ1. Does GPT2SP outperform DL2SP and the baselines for the within-project scenario?}

Table \ref{tbl_within_project} shows the the results we obtained by running GPT2SP, DL2SP, and the Mean and Median baselines on the dataset used and shared publicly by Choetkiertikul et al. \cite{deep2018} (which is also the dataset used by Fu and Tantithamthavorn in their study \cite{fu2022gpt2sp}).

These results reveal that, for the within-project scenario, GPT2SP outperforms the median baseline and DL2SP in only six cases out of 16. Whereas the results of Wicoxon Ranked-Sum test ($\alpha=0.05$) on the distribution absolute errors (see Table \ref{tbl_within_project_wilcox}) reveal that the improvement is statistically significant in only three cases against median (two with negligible and one with small effect size), and two cases against DL2SP (both with negligible effect size). If we also consider Bonferroni correction (by setting $\alpha=0.016$ for $K=3$ hypothesis) the number of statistically significant cases goes down to two and one for the median baseline and DL2SP, respectively. 
Compared to the mean baseline, GPT2SP achieves a lower MAE in 14 cases, whereas it slightly underperforms in one case (Mule) and performs as good in one case (Usergrid). However, the difference between the absolute errors produced by the two methods is statistically significant in 12 cases even when considering Bonferroni correction.

Over all 16 cases and among all four estimation methods, the median baseline achieves the lowest MAE in 8 (50\%) cases, GPT2SP in 5 (31\%) cases, and DL2SP in remaining 3 (19\%) cases.  Therefore, our replication disproves the  results previously obtained by Fu and Tantithamthavorn \cite{fu2022gpt2sp}, that is GPT2SP outperforms DL2SP or the median baseline statistically significantly in all cases for the within project scenario.

\begin{table}[tbh]
\caption{RQ1. Results obtained by GPT2SP, DL2SP, and the two baselines (Mean and Median) for the Within-Project SP scenario. Best results (among all approaches per project) are highlighted in bold.}
\label{tbl_within_project}
\centering
\resizebox{\textwidth}{!}{
\begin{tabular}{l l r r r l l r r r}
\toprule
Project&Method&MAE&MdAE&SA&Project&Method&MAE&MdAE&SA\\
\midrule
Mesos&GPT2SP&1.21&0.98&42.56&Duracloud&GPT2SP&\textbf{0.80}&\textbf{0.50}&\textbf{47.97}\\
&DL2SP&\textbf{1.12}&\textbf{0.73}&\textbf{46.72}&&DL2SP&0.82&0.53&46.97\\
&Mean&1.41&1.78&33.18&&Mean&1.00&1.14&35.17\\
&Median&1.22&2.00&42.30&&Median&0.82&1.00&46.78\\
\midrule
Usergrid &GPT2SP&1.19&1.02&22.56&Data Management&GPT2SP&\textbf{5.39}&\textbf{2.00}&\textbf{54.82}\\
&DL2SP&1.18&\textbf{0.80}&23.25&&DL2SP&5.86&2.22&50.87\\
&Mean&1.19&1.23&22.43&&Mean&8.66&4.55&27.35\\
&Median&\textbf{1.15}&1.00&\textbf{24.97}&&Median&6.19&3.00&48.09\\
\midrule
Appcelerator Studio&GPT2SP&1.53&0.71&49.88&Moodle&GPT2SP&8.38&8.53&45.76\\
&DL2SP&1.42&\textbf{0.58}&53.49&&DL2SP&7.89&\textbf{4.93}&48.92\\
&Mean&1.91&1.52&37.52&&Mean&12.63&12.11&18.21\\
&Median&\textbf{1.30}&1.00&\textbf{57.41}&&Median&\textbf{6.59}&6.00&\textbf{57.32}\\
\midrule
Aptana Studio&GPT2SP&\textbf{3.52}&3.10&\textbf{37.75}&Mule&GPT2SP&2.61&2.23&28.55\\
&DL2SP&4.14&\textbf{2.52}&26.74&&DL2SP&2.59&\textbf{1.96}&29.14\\
&Mean&3.59&3.46&36.53&&Mean&2.60&2.22&28.72\\
&Median&3.61&4.00&36.18&&Median&\textbf{2.47}&2.00&\textbf{32.29}\\
\midrule
Titanium&GPT2SP&2.35&1.45&48.78&Mulesoft&GPT2SP&3.70&2.62&23.55\\
&DL2SP&2.09&\textbf{1.34}&54.49&&DL2SP&3.67&\textbf{2.26}&24.12\\
&Mean&3.02&1.97&34.29&&Mean&3.74&2.80&22.65\\
&Median&\textbf{2.04}&2.00&\textbf{55.71}&&Median&\textbf{3.66}&3.00&\textbf{24.32}\\
\midrule
Bamboo&GPT2SP&0.77&0.69&54.30&Spring XD&GPT2SP&1.78&1.54&37.09\\
&DL2SP&0.81&\textbf{0.61}&51.78&&DL2SP&\textbf{1.70}&\textbf{1.31}&\textbf{39.96}\\
&Mean&1.22&1.31&27.99&&Mean&2.05&2.53&27.59\\
&Median&\textbf{0.75}&1.00&\textbf{55.34}&&Median&1.71&2.00&39.55\\
\midrule
Clover&GPT2SP&3.76&0.97&32.89&Talend Data Quality&GPT2SP&3.65&3.44&29.41\\
&DL2SP&\textbf{3.39}&\textbf{0.80}&\textbf{39.41}&&DL2SP&3.61&\textbf{2.92}&30.07\\
&Mean&4.57&3.06&18.46&&Mean&4.56&5.08&11.69\\
&Median&3.71&2.00&33.77&&Median&\textbf{3.31}&4.00&\textbf{35.85}\\
\midrule
Jira Software&GPT2SP&\textbf{1.57}&\textbf{0.81}&\textbf{54.09}&Talend ESB&GPT2SP&\textbf{0.86}&\textbf{0.55}&\textbf{38.81}\\
&DL2SP&1.70&1.09&50.08&&DL2SP&0.90&0.59&36.19\\
&Mean&2.40&2.15&29.61&&Mean&1.04&0.91&26.47\\
&Median&2.31&2.00&32.40&&Median&0.92&1.00&34.86\\
\bottomrule
\end{tabular}
}
\end{table}

\begin{table}[tbh]
\caption{RQ1. Results of the Wilcoxon test (with Vargha-Delaney $\hat{A}_{12}$ effect size in parentheses) comparing the absolute errors of GPT2SP to that of DL2SP and the two baselines (i.e., Mean and Median).}
\label{tbl_within_project_wilcox}
\centering
\resizebox{0.6\textwidth}{!}{
\begin{tabular}{l r r r }
\toprule
\multirow{2}{*}{Project}&\multicolumn{3}{l}{GPT2SP vs}\\\cmidrule{2-4}
&Mean&Median&DL2SP\\
\midrule
Mesos&0.127 (0.55)&0.955 (0.43)&0.248 (0.53)\\
Usergrid &\textless 0.001 (0.70)&1.000 (0.43)&0.999 (0.44)\\
Appcelerator Studio&\textless 0.001 (0.64)&1.000 (0.37)&1.000 (0.43)\\
Aptana Studio&0.713 (0.48)&0.679 (0.49)&0.044 (0.55)\\
Titanium&\textless 0.001 (0.70)&0.583 (0.49)&0.516 (0.50)\\
Bamboo&\textless 0.001 (0.70)&0.021 (0.59)&0.412 (0.51)\\
Clover&\textless 0.001 (0.76)&0.004 (0.54)&0.001 (0.54)\\
Jira Software&\textless 0.001 (0.62)&0.988 (0.42)&0.478 (0.50)\\
Duracloud&\textless 0.001 (0.58)&0.663 (0.49)&0.972 (0.46)\\
Data Management&\textless 0.001 (0.79)&1.000 (0.31)&1.000 (0.40)\\
Moodle&0.323 (0.51)&0.678 (0.49)&0.294 (0.52)\\
Mule&0.497 (0.50)&0.751 (0.48)&0.722 (0.48)\\
Mulesoft&\textless 0.001 (0.59)&0.962 (0.47)&0.843 (0.48)\\
Spring XD&\textless 0.001 (0.66)&0.977 (0.45)&0.433 (0.50)\\
Talend Data Quality&\textless 0.001 (0.61)&0.130 (0.53)&0.325 (0.51)\\
Talend ESB&\textless 0.001 (0.68)&\textless 0.001 (0.65)&0.217 (0.54)\\
\bottomrule
\end{tabular}
}
\end{table}

\textbf{RQ2. Does GPT2SP outperform DL2SP and the baselines for the cross-project scenario?}

Similar to Choetkiertikul et al. \cite{deep2018} and Tawosi et al. \cite{tawosi2022deep}, Fu and Tantithamthavorn also investigated two cross-project estimation scenarios, namely the within-repository and cross-repository scenarios.

Table \ref{tbl.within-repo} shows the results we achieved for GPT2SP, DL2SP and the mean and median baselines for the within-repository scenario. The results of the Wilcoxon test on the distribution of the absolute errors produced by GPT2SP against each of the other methods is provided in the last column of this table.

We can observe that GPT2SP outperforms DL2SP in only two out of eight cases  (with marginal statistical significance and a negligible effect size in only one of the two cases), whereas DL2SP outperforms GPT2SP in six cases.

GPT2SP outperforms the Median baseline in only two cases out of eight, with statistical significance but negligible effect size in both cases. GPT2SP achieves a lower MAE than that of the Mean baseline in five cases, a higher one in two, and equal one in the remaining one case out of eight cases, where the difference is significant for three cases but always with a negligible effect size.

When considering the results of cross-repository scenario in Table \ref{tbl.cross-repo}, we can observe that GPT2SP outperforms DL2SP in six cases out of eight with statistically significant difference in five of them out of which one show a medium effect size, one a small one, and three a negligible one.
GPT2SP achieves a lower MAE than that obtained by Median in three cases only, among which one case shows a small effect size and two other cases show a negligible effect size. GPT2SP outperforms the Mean baseline in seven cases with statistically significant difference in six with a large effect size in three, medium one in two and small in one.

Overall, considering both cross-project scenarios, GPT2SP performs better than DL2SP in the cross-repository scenario (the same conclusion achieved by the original study).
However, GPT2SP performs poorly when compared against the na\"ive Median baseline, that is, it achieves statistically significantly better results in less than a third of the cases in both scenarios combined (i.e., five out of 16 cases). Therefore, the replication results does not support the conclusion made in the original study stating that GPT2SP outperforms the baselines statistically significantly.

\begin{table}[]
\caption{RQ2. Comparing GPT2SP cross-project SP estimation replication results to DL2SP \cite{deep2018}, and to the baselines. The results of the Wilcoxon test ($\hat{A}_{12}$ effect size in parentheses) for GPT2SP vs. DL2SP and baselines are shown in the last column. Best results (among all approaches per project) are highlighted in bold.}
\label{tbl.cross-project}
\centering
\begin{subtable}[t]{0.49\textwidth}
    \centering
    \caption{Within-Repository Training}
    \label{tbl.within-repo}
    \resizebox{\textwidth}{!}{
    \begin{tabular}{l l l r r r r}
    \toprule
    Source&Target&Method&MAE&MdAE&SA&GPT2SP vs.\\
\midrule
MESOS&USERGRID&GPT2SP&1.11&0.84&42.49&\\
(ME)&(UG)&DL2SP&1.16&0.96&39.84&0.049 (0.53) \\
&&Mean&1.02&0.19&46.99&0.998 (0.45) \\
&&Median&\textbf{0.89}&\textbf{0.00}&\textbf{54.10}&1.000 (0.37) \\
\midrule
USERGRID&MESOS&GPT2SP&1.52&0.92&15.64&\\
(UG)&(ME)&DL2SP&1.51&1.01&16.18&0.662 (0.50) \\
&&Mean&1.52&\textbf{0.80}&15.57&0.325 (0.50) \\
&&Median&\textbf{1.50}&1.00&\textbf{16.27}&0.692 (0.50) \\
\midrule
TISTUD&APSTUD&GPT2SP&4.62&3.37&7.99&\\
(AS)&(AP)&DL2SP&4.37&2.98&12.99&1.000 (0.45) \\
&&Mean&\textbf{4.27}&\textbf{2.18}&\textbf{15.05}&0.992 (0.47) \\
&&Median&4.38&3.00&12.73&1.000 (0.43) \\
\midrule
TISTUD&TIMOB&GPT2SP&3.28&2.47&22.60&\\
(AS)&(TI)&DL2SP&3.38&2.39&20.31&0.253 (0.51) \\
&&Mean&3.45&2.82&18.69&\textless 0.001 (0.56) \\
&&Median&\textbf{3.17}&\textbf{2.00}&\textbf{25.24}&1.000 (0.44) \\
\midrule
APSTUD&TISTUD&GPT2SP&2.86&2.45&45.15&\\
(AP)&(AS)&DL2SP&\textbf{2.70}&\textbf{2.07}&\textbf{48.25}&1.000 (0.47) \\
&&Mean&3.38&3.24&35.20&\textless 0.001 (0.58) \\
&&Median&3.17&3.00&39.30&\textless 0.001 (0.55) \\
\midrule
APSTUD&TIMOB&GPT2SP&4.09&3.52&30.38&\\
(AP)&(TI)&DL2SP&\textbf{3.51}&\textbf{2.53}&\textbf{40.34}&1.000 (0.39) \\
&&Mean&4.36&4.24&25.78&\textless 0.001 (0.56) \\
&&Median&4.19&4.00&28.67&0.004 (0.52) \\
\midrule
MULE&MULESTUDIO&GPT2SP&3.48&2.29&21.32&\\
(MU)&(MS)&DL2SP&3.64&\textbf{2.04}&17.61&0.206 (0.51) \\
&&Mean&3.34&2.71&24.42&0.943 (0.48) \\
&&Median&\textbf{3.26}&3.00&\textbf{26.26}&0.999 (0.45) \\
\midrule
MULESTUDIO&MULE&GPT2SP&3.03&2.65&30.27&\\
(MS)&(MU)&DL2SP&2.77&2.47&36.28&0.985 (0.47) \\
&&Mean&3.05&\textbf{1.77}&29.83&0.078 (0.52) \\
&&Median&\textbf{2.60}&3.00&\textbf{40.24}&1.000 (0.43) \\
    \bottomrule
    \end{tabular}
    }
\end{subtable}%
\hspace{0.49cm}
\begin{subtable}[t]{0.47\textwidth}
    \centering
    \caption{Cross-Repository Training}
    \label{tbl.cross-repo}
    \resizebox{\textwidth}{!}{
    \begin{tabular}{l l l r r r r}
    \toprule
    Source&Target&Method&MAE&MdAE&SA&GPT2SP vs.\\
\midrule
TISTUD&USERGRID&GPT2SP&\textbf{2.18}&2.03&\textbf{37.67}&\\
(AS)&(UG)&DL2SP&3.47&3.50&0.98&\textless 0.001 (0.76) \\
&&Mean&3.08&2.82&12.02&\textless 0.001 (0.73) \\
&&Median&2.30&\textbf{2.00}&34.40&\textless 0.001 (0.56) \\
\midrule
TISTUD&MESOS&GPT2SP&2.65&\textbf{2.75}&28.83&\\
(AS)&(ME)&DL2SP&3.18&3.20&14.70&\textless 0.001 (0.60) \\
&&Mean&3.28&2.82&11.95&\textless 0.001 (0.64) \\
&&Median&\textbf{2.58}&3.00&\textbf{30.85}&0.994 (0.47) \\
\midrule
MDL&APSTUD&GPT2SP&4.37&3.07&68.12&\\
(MD)&(AP)&DL2SP&5.03&3.77&63.29&\textless 0.001 (0.55) \\
&&Mean&9.84&8.95&28.21&\textless 0.001 (0.86) \\
&&Median&\textbf{3.97}&\textbf{3.00}&\textbf{71.05}&0.989 (0.47) \\
\midrule
MDL&TIMOB&GPT2SP&3.60&2.53&73.90&\\
(MD)&(TI)&DL2SP&\textbf{3.34}&\textbf{1.96}&\textbf{75.82}&1.000 (0.45) \\
&&Mean&11.19&11.95&18.91&\textless 0.001 (0.91) \\
&&Median&4.19&4.00&69.63&\textless 0.001 (0.60) \\
\midrule
MDL&TISTUD&GPT2SP&\textbf{2.16}&1.96&\textbf{83.92}&\\
(MD)&(AS)&DL2SP&2.64&\textbf{1.66}&80.35&\textless 0.001 (0.57) \\
&&Mean&11.45&11.95&14.90&\textless 0.001 (0.98) \\
&&Median&3.17&3.00&76.44&\textless 0.001 (0.61) \\
\midrule
DM&TIMOB&GPT2SP&3.58&2.31&61.99&\\
&(TI)&DL2SP&3.81&2.65&59.52&\textless 0.001 (0.53) \\
&&Mean&5.61&5.03&40.35&\textless 0.001 (0.75) \\
&&Median&\textbf{3.46}&\textbf{1.00}&\textbf{63.22}&0.999 (0.47) \\
\midrule
USERGRID&MULESTUDIO&GPT2SP&4.02&2.16&4.57&\\
(UG)&(MS)&DL2SP&3.95&2.11&6.18&0.865 (0.48) \\
&&Mean&4.04&2.20&4.00&0.145 (0.52) \\
&&Median&\textbf{3.91}&\textbf{2.00}&\textbf{7.16}&0.996 (0.46) \\
\midrule
MESOS&MULE&GPT2SP&3.13&2.24&9.57&\\
(ME)&(MU)&DL2SP&3.20&2.31&7.37&0.347 (0.51) \\
&&Mean&\textbf{2.89}&\textbf{1.81}&\textbf{16.56}&0.999 (0.46) \\
&&Median&2.92&2.00&15.65&0.999 (0.46) \\
    \bottomrule
    \end{tabular}
    }
\end{subtable}
\end{table}

\bibliographystyle{IEEEtran}
\bibliography{gpt2sp_replication_report}

\begin{thebibliography}{1}
\providecommand{\url}[1]{#1}
\csname url@samestyle\endcsname
\providecommand{\newblock}{\relax}
\providecommand{\bibinfo}[2]{#2}
\providecommand{\BIBentrySTDinterwordspacing}{\spaceskip=0pt\relax}
\providecommand{\BIBentryALTinterwordstretchfactor}{4}
\providecommand{\BIBentryALTinterwordspacing}{\spaceskip=\fontdimen2\font plus
\BIBentryALTinterwordstretchfactor\fontdimen3\font minus
  \fontdimen4\font\relax}
\providecommand{\BIBforeignlanguage}[2]{{%
\expandafter\ifx\csname l@#1\endcsname\relax
\typeout{** WARNING: IEEEtran.bst: No hyphenation pattern has been}%
\typeout{** loaded for the language `#1'. Using the pattern for}%
\typeout{** the default language instead.}%
\else
\language=\csname l@#1\endcsname
\fi
#2}}
\providecommand{\BIBdecl}{\relax}
\BIBdecl

\bibitem{fu2022gpt2sp}
M.~Fu and C.~Tantithamthavorn, ``{GPT2SP}: A transformer-based agile story
  point estimation approach,'' \emph{IEEE Transactions on Software
  Engineering}, 2022.

\bibitem{deep2018}
M.~Choetkiertikul, H.~K. Dam, T.~Tran, T.~Pham, A.~Ghose, and T.~Menzies, ``A
  deep learning model for estimating story points,'' \emph{IEEE TSE}, vol.~45,
  no.~7, pp. 637--656, 2019.

\bibitem{tawosi2022deep}
V.~Tawosi, R.~Moussa, and F.~Sarro, ``Deep learning for agile effort estimation
  have we solved the problem yet?'' \emph{arXiv preprint arXiv:2201.05401},
  2022.

\end{thebibliography}

\end{document}